\documentstyle[sprocl,psfig]{article}

\def\be{\begin{equation}}
\def\ee{\end{equation}}
\def\bea{\begin{eqnarray}}
\def\eea{\end{eqnarray}}
\newcommand{\lsim}{\mathrel{\mathop{\kern 0pt \rlap
  {\raise.2ex\hbox{$<$}}}
  \lower.9ex\hbox{\kern-.190em $\sim$}}}
\newcommand{\gsim}{\mathrel{\mathop{\kern 0pt \rlap
  {\raise.2ex\hbox{$>$}}}
  \lower.9ex\hbox{\kern-.190em $\sim$}}}

\begin{document}

\thispagestyle{empty}

\rightline{October 1998}
\rightline{DFTT 63/98}

\vspace{1pc}
\centerline{\Large \bf SUPERSYMMETRIC DARK MATTER} 
\smallskip
\centerline{\Large \bf MSSM and SUGRA schemes in the light of a possible annual}
\smallskip
\centerline{\Large \bf  modulation effect in WIMP direct search
\footnote{Report on the work done in collaboration with A. Bottino, F. Donato and S. Scopel.}}

\vspace{3pc}
\centerline{\large \bf Nicolao Fornengo}

\vspace{1pc}
\em
\begin{center}
\begin{tabular}{c}
Dipartimento di Fisica Teorica, Universit\`a di Torino \\
and \\
INFN, Sezione di Torino \\
Via P. Giuria 1, 10125 Torino, Italy
\\
{\sl fornengo@to.infn.it}
\\
{\sl http://www.to.infn.it/$\tilde{\hspace{1ex}}$fornengo/index.html}
\end{tabular}
\end{center}

\vspace{2pc}
\rm
\centerline{\large \bf Abstract}
\bigskip

Recently the DAMA/NaI Collaboration reported further indication of a possible
modulation effect in WIMP direct detection. In this note we discuss the
relevance of this result for supersymmetric theories where the neutralino
is the dark matter candidate. We specifically consider the Minimal Supersymmetric
extension of the Standard Model (MSSM) and Supergravity-inspired schemes, with
possible deviations of the unification conditions at the GUT scale in the
Higgs sector. The main results of our analysis show that the annual modulation 
data are widely compatible with an interpretation in terms of a relic neutralino 
as the major component of dark matter in the Universe. We also discuss the
implications for searches of supersymmetry at accelerators and for indirect 
searches of neutralino dark matter. 
\vspace{1pc}

\vfill

\begin{center}
{ \em
Talk presented by Nicolao Fornengo at the 2 $^{\rm nd}$
 ``International Workshop on the Identification of Dark
  Matter (IDM'98)", Buxton, England} \\
{\em September 7--11, 1998}
\end{center}

\vspace{2pc}
\eject



\setcounter{page}{1}

\title{SUPERSYMMETRIC DARK MATTER \\ MSSM and SUGRA schemes in the
light of a possible annual modulation effect in WIMP direct search
\footnote[1]{Report on the work done in collaboration with A. Bottino, F. Donato and S. Scopel.}}

\author{Nicolao FORNENGO}

\address{Dipartimento di Fisica Teorica, Universit\`a di Torino \\
and INFN - Sezione di Torino,
via P. Giuria 1, 10125 Torino, Italy \\
E-mail: fornengo@to.infn.it}  

\maketitle
\abstracts{
Recently the DAMA/NaI Collaboration reported further indication of a possible
modulation effect in WIMP direct detection. In this note we discuss the
relevance of this result for supersymmetric theories where the neutralino
is the dark matter candidate. We specifically consider the Minimal Supersymmetric
extension of the Standard Model (MSSM) and Supergravity-inspired schemes, with
possible deviations of the unification conditions at the GUT scale in the
Higgs sector. The main results of our analysis show that the annual modulation 
data are widely compatible with an interpretation in terms of a relic neutralino 
as the major component of dark matter in the Universe. We also discuss the
implications for searches of supersymmetry at accelerators and for indirect 
searches of neutralino dark matter. 
}

\vspace{-8.0mm}
\section{Introduction}

\vspace{-3.0mm}
The presence of non--baryonic dark matter in our Galaxy can 
be probed by means of different techniques which attempt to
detect either directly (through elastic scattering off
nuclei) or indirectly (through products of annihilation)
the dark matter particles which are supposed to be embedded 
in the galactic halo. It has been shown (see, e.g., 
Ref. \cite{ringberg98}) that the sensitivities of the
present experiments is currently at the level required 
for the study of one of the most appealing particle
candidates for dark matter, the neutralino. This implies
that it is now feasible to start the investigation of possible
signatures in the detection rates, originated
by specific features related to the presence of the dark matter
particles. 

In the case of direct detection, a typical signature consists 
in the annual modulation of the detection rate. 
This effect was first pointed out in the seminal papers of 
Refs. \cite{ann_mod_th}, where it 
was observed that, during the orbital motion of the Earth around the Sun,
the change of direction of the relic particle velocities
with respect to the detector induces a time dependence in the 
differential detection rate, i.e.
$S(E,t) = S_0 (E) + S_m (E) \cos [\omega (t-t_0)]$,
where $\omega = 2\pi/365$ days and $t_0 = 153$ days
(June 2$^{\rm nd}$). $S_0 (E)$ is the average
(unmodulated) differential rate and $S_m (E)$ is the
modulation amplitude of the rate.
The relative importance of $S_m (E)$ with respect to $S_0 (E)$ for a given
detector, depends both on the mass of the dark matter particle and on the value of the
recoil energy where the effect is looked at. Typical values of $S_m (E)/S_0 (E)$
for a NaI detector range from a few percent up to $\sim$ 15\%,
for WIMP masses of the order of 20--80 GeV and recoil energies
below 8--10 KeV. 

Over the last year, the DAMA/NaI Collaboration reported on two
different analyses of the data collected during two periods
of data taking, obtained with an experimental set--up
consisting of nine 9.70 Kg NaI(Tl) detectors \cite{dama1,dama2}.
The data have been analysed by employing a maximum likelihood
method, which allows to test the hypothesis of the presence
of a yearly modulated signal against a time--independent background,
by properly considering the energy and time behaviour expected 
for a recoil of massive WIMPs. The remarkable result of 
Refs. \cite{dama1,dama2} is that the data support the possible presence of an 
annual modulation effect induced by WIMPS in the counting rate of the detector. 
By combining the data of the two periods of data taking, a total
exposure of 19,511 kg $\times$ day has been collected. The maximum
likelihood analysis indicates that the hypothesis of
presence of modulation against the hypothesis of absence of  modulation
is statistically favoured at 99.6\% C.L., and pins
down a 2--$\sigma$ C.L. region in the plane 
$\xi \sigma^{(\rm nucleon)}_{\rm scalar}$ -- $m_\chi$, 
where $m_\chi$ is the WIMP mass, 
$\sigma^{(\rm nucleon)}_{\rm scalar}$ is the WIMP--nucleon  scalar elastic
cross section  and $\xi = \rho_\chi / \rho_l$ 
is the fractional amount of local 
WIMP density $\rho_\chi$ with respect to the total local 
dark matter density $\rho_l$. This region is plotted in
Fig.1 as a closed dashed curve, for $\rho_l=0.3$ GeV cm$^{-3}$.
The ensuing 1--$\sigma$ ranges for the two quantities 
are: $m_{\chi} = 59_{- 14}^{+ 22}$ GeV and 
$\xi \sigma^{(\rm nucleon)}_{\rm scalar} = 7.0_{-1.7}^{+0.4} \times 10^{-9}$ nb \cite{dama2}. 
These results refer to $v_{\rm rms}$ = 270 Km s$^{-1}$ 
for the velocity dispersion of the WIMP Maxwellian 
velocity distribution in the halo, $v_{\rm esc} = 650$ Km s$^{-1}$
for the WIMP escape velocity and $v_\odot = 232$ Km s$^{-1}$
for the velocity of the Sun around the galactic centre. We
notice that a variation of these velocity parameters inside their
2--$\sigma$ ranges of uncertainty affects the maximum likelihood
value of $m_\chi$ and $\xi \sigma^{(\rm nucleon)}_{\rm scalar}$
by roughly $\pm$ 30\% and by $\sim \pm$ 12\%, respectively.

In this paper, which is based on the results obtained in 
Refs. \cite{mssm_mod,sugra_mod,ind_mod} we derive the theoretical implications of
the experimental data of Refs. \cite{dama1,dama2}, assuming that the indication of the
possible annual modulation is due to relic
neutralinos and we show that these data are fully compatible with an
interpretation in terms of a relic neutralino as the major component of dark
matter in the Universe. We select the relevant susy
configurations in two different frameworks: the low energy
minimal supersymmetric extension of the standard model (MSSM) and
supergravity--inspired schemes (SUGRA models). In the latter case,
we will explicitly consider both a strict unification at the GUT scale 
and the possibility of deviation from universality in the Higgs sector.
The present analysis extends a previous analysis of ours \cite{pinning} 
referring to the experimental data of Ref. \cite{dama1}.
We then finally also discuss how the susy configurations, selected by 
the annual modulation data, can be investigated by
indirect searches for relic WIMPs and at accelerators. 

\vspace{-6.0mm}
\section{Selection of susy configurations by the annual modulation data}

\begin{figure}[t]
\hbox{
\hspace{-3.5mm}
\psfig{figure=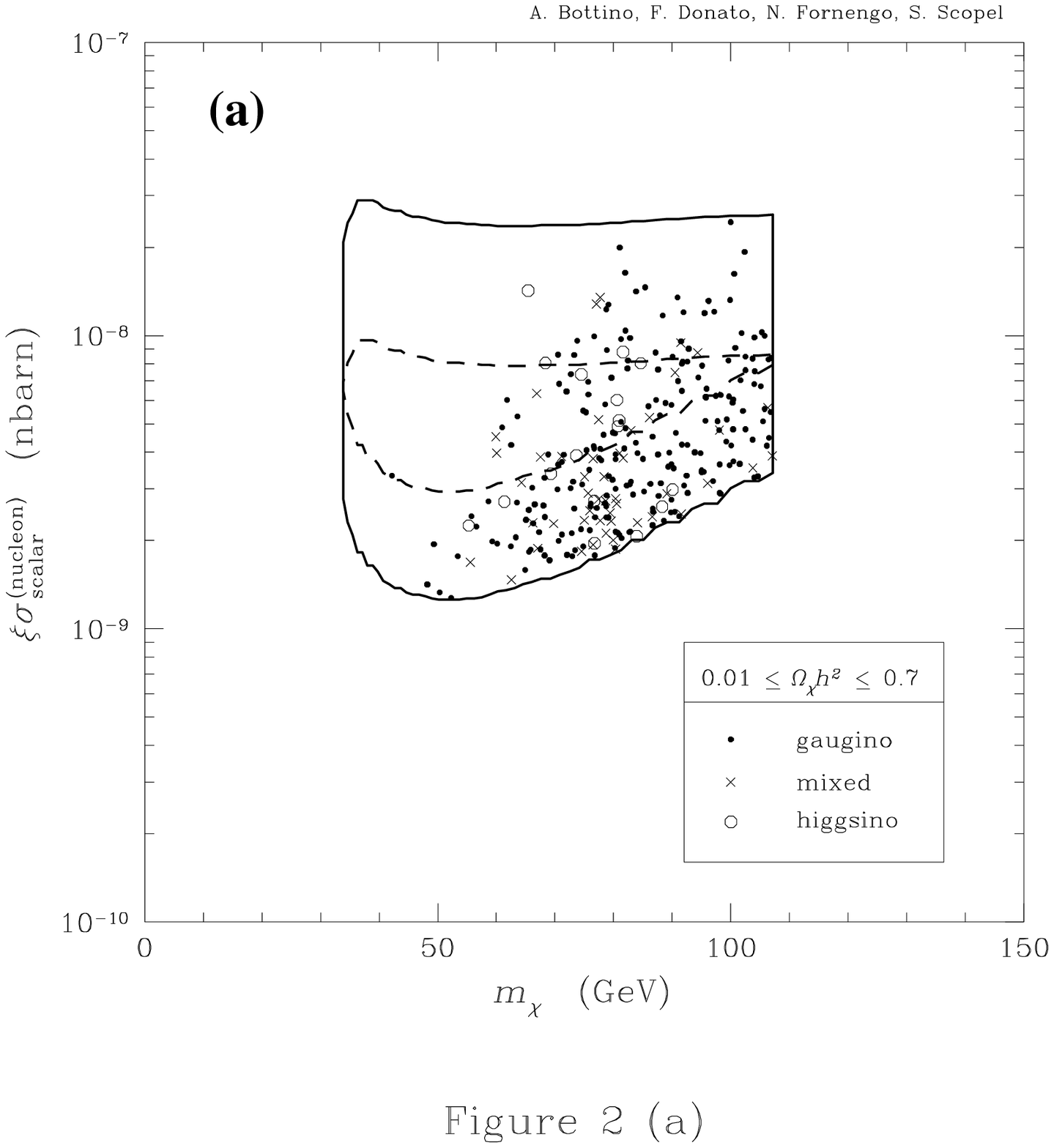,width=2.45in,bbllx=60bp,bblly=222bp,bburx=530bp,bbury=675bp,clip=}
\hspace{-4.5mm}
\psfig{figure=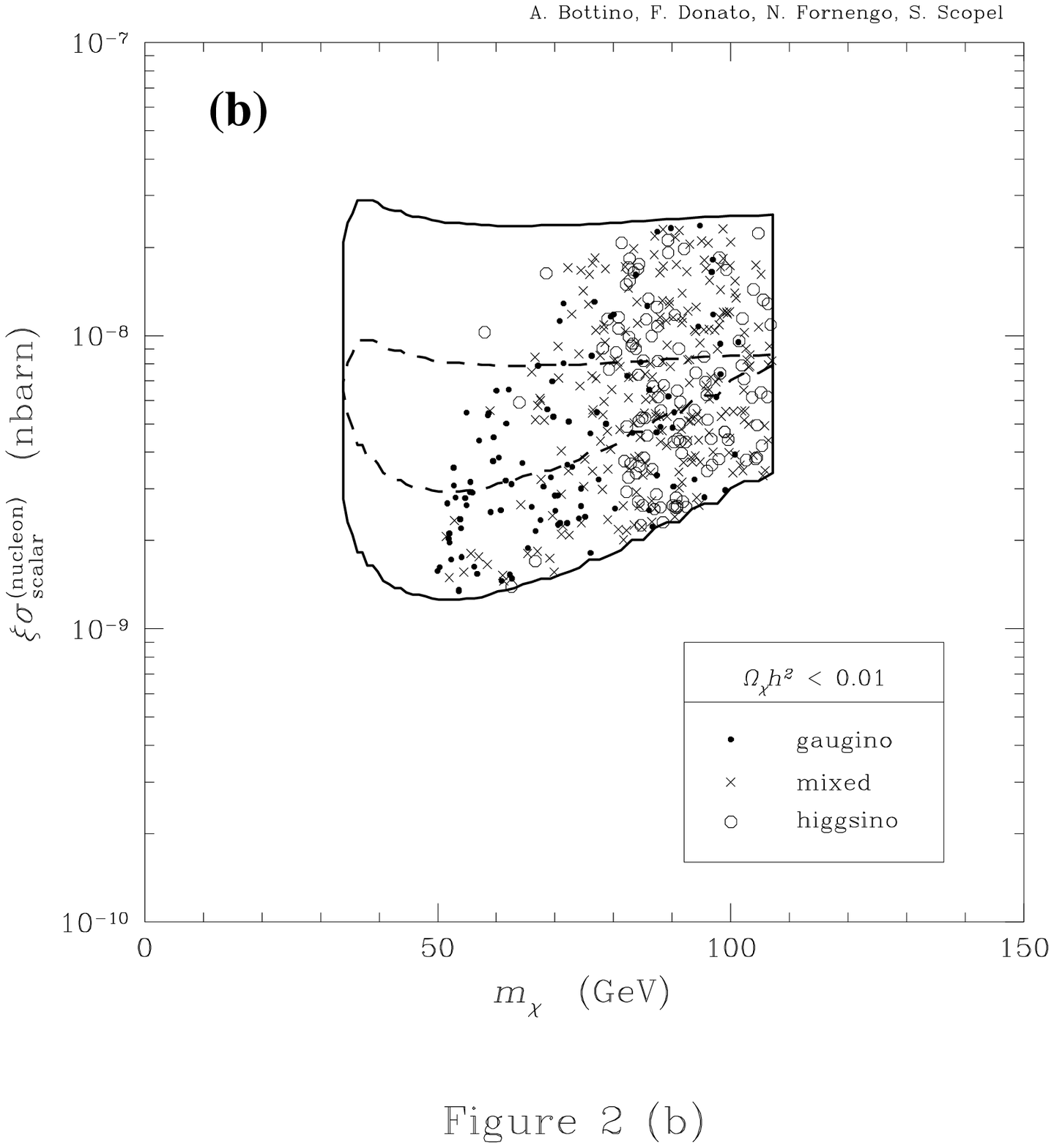,width=2.45in,bbllx=60bp,bblly=222bp,bburx=530bp,bbury=675bp,clip=}
}
\vspace{-3.0mm}
\caption{Scatter plot of {\bf MSSM} configurations compatible with the annual
modulation data in the plane 
$m_{\chi}$--$\xi \sigma_{\rm scalar}^{(\rm nucleon)}$. 
The dashed contour line
delimits the 2--$\sigma$ C.L. region, obtained by the DAMA/NaI Collaboration, 
by combining together the data of
the two running periods of the annual modulation experiment.  
The solid contour line is obtained from the dashed line, which refers to 
$\rho_l = 0.3$ GeV cm$^{-3}$, by accounting for 
the uncertainty range of $\rho_l$.
(a) and (b) refer to configurations with 
$0.01 \leq \Omega_{\chi} h^2 \leq 0.7$ and with $\Omega_{\chi} h^2 < 0.01$, 
respectively.}
\vspace{-4.0mm}
\end{figure}

\vspace{-3.0mm}
In order to discuss which region in the susy parameter space is selected by the 
DAMA data, we first convert the region delimited by the 2--$\sigma$ 
C.L. dashed contour line of Fig.1 
into an enlarged one, which accounts for the uncertainty in the value of 
$\rho_l$, due to a possible flattening of the dark matter
halo and a possibly sizeable baryonic contribution to the
galactic dark matter:
0.1 GeV cm$^{-3} \leq \rho_l \leq $ 0.7 GeV cm$^{-3}$. One then obtains the
2--$\sigma$ C.L. region denoted by a solid contour in Fig.1
(hereafter denoted as region $R$).
The susy configurations are then selected by the requirement that 
($m_{\chi}, \xi \sigma_{\rm scalar}^{(\rm nucleon)}) \in R$, when
$m_{\chi}$, $\sigma_{\rm scalar}^{(\rm nucleon)}$ and $\xi$ are
evaluated in the MSSM and SUGRA schemes.
As for the values to be assigned to the quantity $\xi = \rho_{\chi}/ \rho_l$ 
we adopt the standard rescaling recipe \cite{mssm_mod}: 
$\xi = {\rm min} [1, {\Omega_\chi h^2 / (\Omega h^2)_{\rm min}}]$,
where $\Omega_\chi h^2$ denotes the neutralino relic abundance,
calculated in the susy model \cite{omega}, and $(\Omega h^2)_{\rm min}$
is a minimal value compatible with observational data and with 
large--scale structure calculations. We use here the value 
$(\Omega h^2)_{\rm min} = 0.01$ \cite{mssm_mod}.
In all our analyses, we consider as cosmologically acceptable 
all configurations which provide $\Omega_\chi h^2 \leq 0.7$.

\vspace{-5.0mm}
\section{Analysis in the MSSM}

\begin{figure}[t]
\centerline{
\psfig{figure=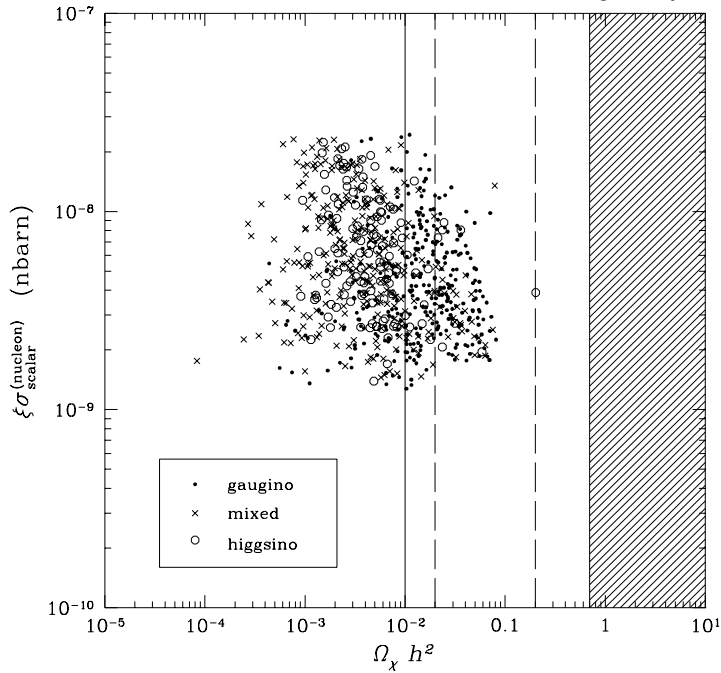,width=4.00in,bbllx=40bp,bblly=190bp,bburx=530bp,bbury=650bp,clip=}
}
\vspace{-5.0mm}
\caption{Scatter plot of {\bf MSSM} configurations compatible with the annual
modulation data in the plane 
 $\Omega_{\chi} h^2$ -- $\xi \sigma_{\rm scalar}^{(\rm nucleon)}$.
The two vertical solid lines 
delimit the $\Omega_{\chi} h^2$--range of cosmological interest. The two
dashed lines delimit the most appealing interval for 
$\Omega_{\chi} h^2$, as suggested by the most recent observational data. 
The hatched area is excluded by cosmology.}
\vspace{-4.0mm}
\end{figure}

\begin{figure}[t]
\hbox{
\hspace{-3.5mm}
\psfig{figure=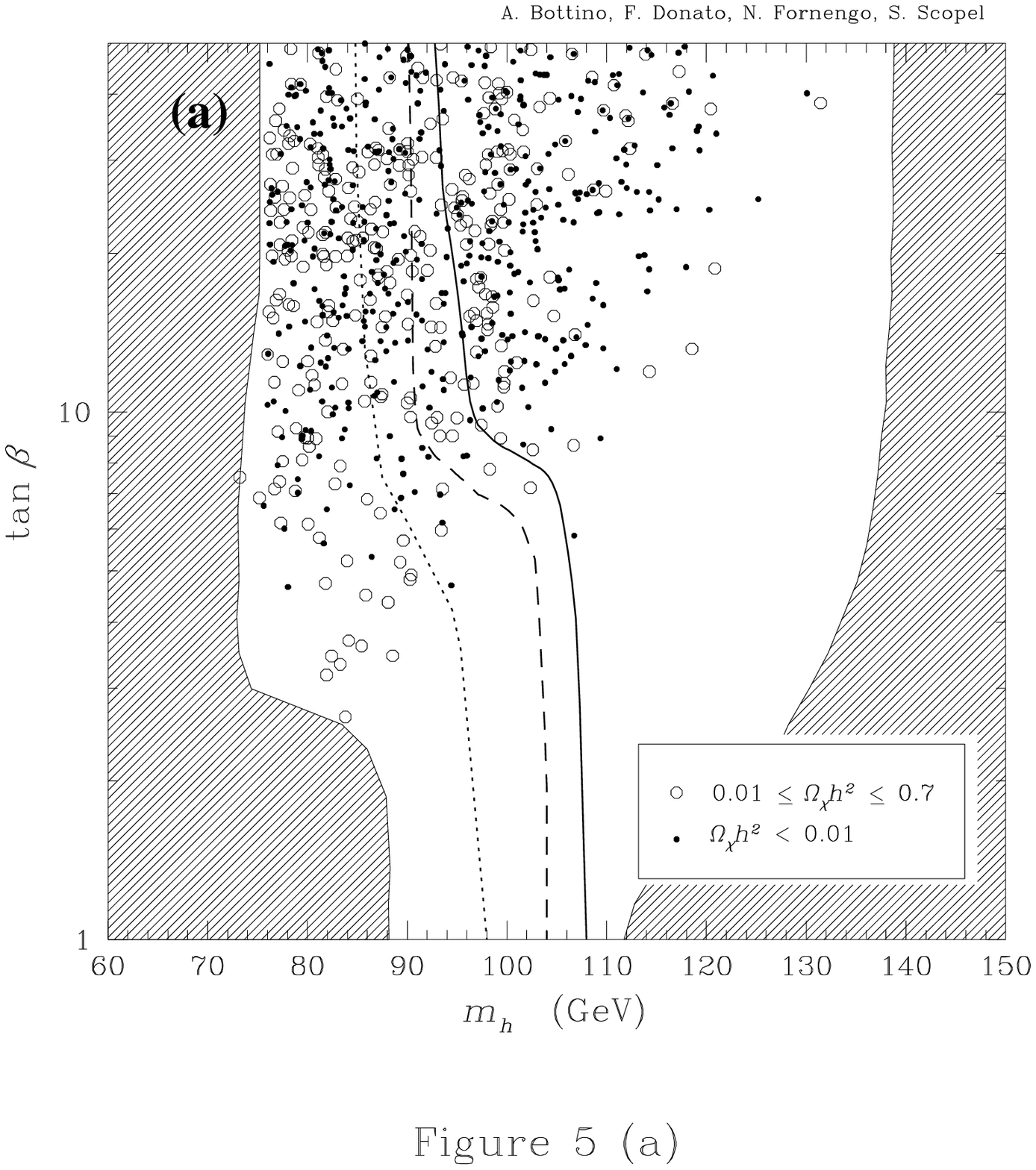,width=2.45in,bbllx=60bp,bblly=222bp,bburx=530bp,bbury=675bp,clip=}
\hspace{-4.5mm}
\psfig{figure=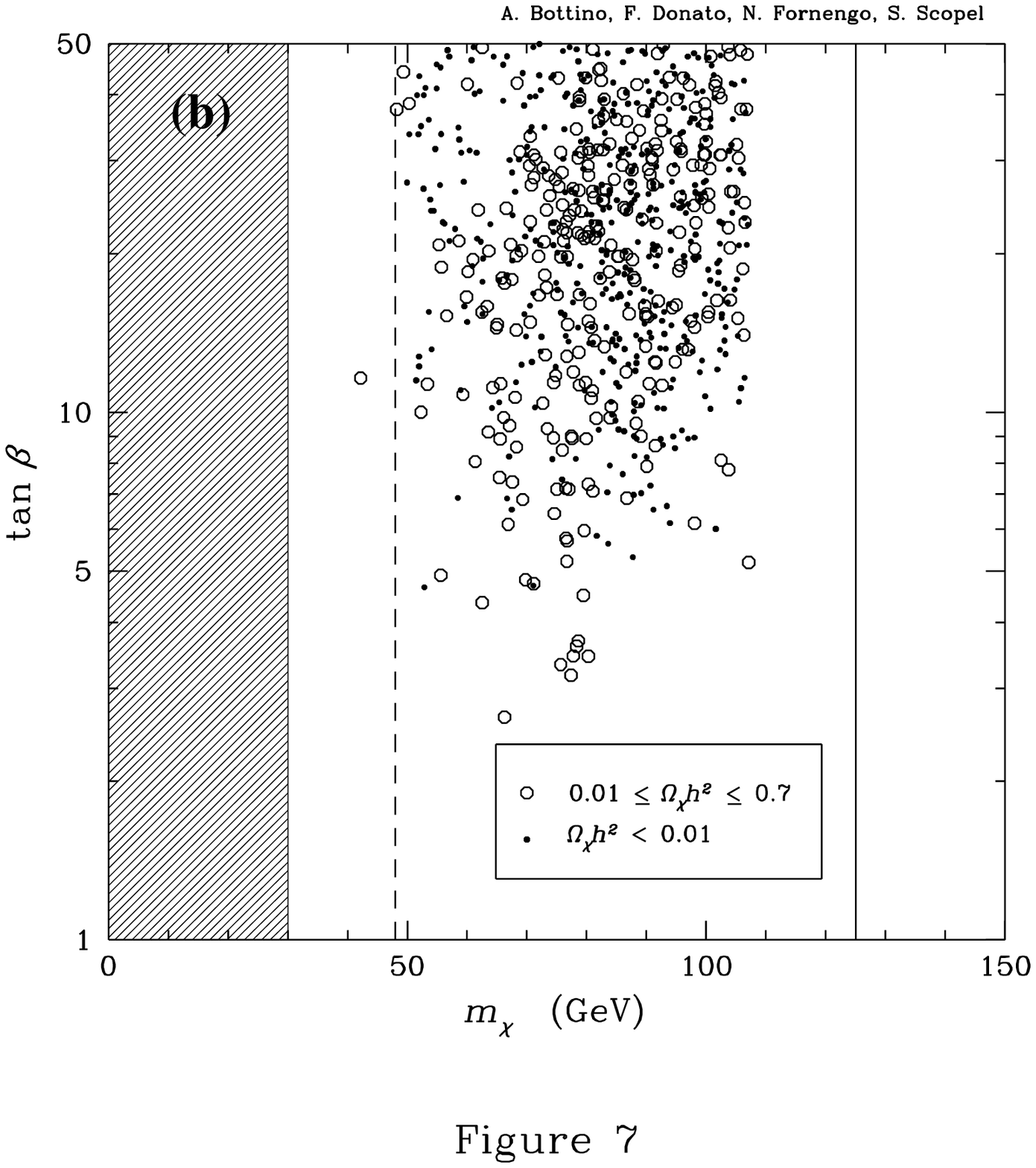,width=2.45in,bbllx=60bp,bblly=222bp,bburx=530bp,bbury=675bp,clip=}
}
\vspace{-3.0mm}
\caption{Explorability at accelerators of the {\bf MSSM} configurations compatible with the annual
modulation data. (a): Scatter plot in the plane $m_h$ -- $\tan \beta$. 
The hatched region on the right is excluded by theory. 
The hatched region on the left is 
excluded by present LEP data at 183 GeV. The dotted and the dashed 
curves denote the reach of LEP2 at energies 192 GeV and 
200 GeV, respectively. The solid line represents the 
95\% C.L. bound reachable at LEP2, in case of non discovery of a neutral 
Higgs boson. (b): Scatter plot in the plane 
$m_{\chi}$ -- $\tan \beta$. The hatched region on the left is 
excluded by present LEP data. The dashed and the 
solid vertical lines denote the reach of LEP2 and TeV33, 
respectively.}
\vspace{-4.0mm}
\end{figure}

\vspace{-3.0mm}
In our first analysis \cite{mssm_mod} we employ the 
minimal supersymmetric extension of the standard model (MSSM)
which conveniently  describes the 
susy phenomenology at the electroweak scale, without too strong 
theoretical assumptions.
The MSSM is based on the same gauge group as the Standard Model, 
contains the susy extension of its particle content and 
two Higgs doublets. The free parameters of the
model are: the SU(2) gaugino mass parameter $M_2$, related to 
the U(1) gaugino mass parameter by the GUT relation
$M_1= (5/3) \tan^2 \theta_W M_2$; the Higgs--mixing 
parameter $\mu$; the ratio of the two Higgs
vev's $\tan\beta$; the mass of the pseudoscalar neutral Higgs $m_A$;
a common soft--mass parameter for all the squarks and sfermions $m_0$;
a common trilinear parameter for the third family $A$ (the other trilinear parameters
are all set to zero).
The parameters are varied in the following ranges:
$10\;\mbox{GeV} \leq M_2 \leq  500\;\mbox{GeV},\;
10\;\mbox{GeV} \leq |\mu| \leq  500\;\mbox{\rm GeV},\;
75\;\mbox{GeV} \leq m_A \leq  1\;\mbox{TeV},\; 
100\;\mbox{GeV} \leq m_0 \leq  1\;\mbox{TeV},\;
-3 \leq A \leq +3,\;
1 \leq \tan \beta \leq 50$. 

Our susy parameter space is constrained by
the latest data from 
LEP2 on Higgs, neutralino, chargino and 
sfermion masses \cite{LEP}. Moreover, the constraints 
due to the $b \rightarrow s + \gamma$ process 
has been taken into account, considering the 
latest results both in the theoretical evaluation 
and in the experimental determination of the
branching ratio \cite{bsgamma}.

By varying the susy parameters inside the ranges defined above,
we find that a large portion of the modulation region $R$ is indeed covered 
by susy configurations, 
compatible with all present physical constraints. This set of susy states (set $S$)
is displayed in Fig.1 with 
different symbols, depending on the neutralino composition. 
In Fig.1(a) we notice that a quite sizeable portion of region $R$ is 
populated by susy configurations
with neutralino relic abundance inside the cosmologically interesting range 
$0.01 \lsim \Omega_{\chi} h^2 \lsim 0.7$. 
Thus we obtain  the first main result of our analysis, i.e. 
{\it the annual 
modulation region, singled out by the DAMA/NaI experiment, is largely 
compatible with a relic neutralino as the major component of dark matter}. 
This is certainly the most remarkable possibility. However, we also keep under
consideration neutralino configurations with a small contribution to 
$\Omega_{\chi} h^2$ (see Fig.1(b)), since also the detection of relic 
particles with these features would provide in itself a very noticeable 
information. 

The neutralino relic abundance $\Omega_{\chi} h^2$ is plotted versus 
$\xi \sigma_{\rm scalar}^{(\rm nucleon)}$ in 
Fig.2. We notice that a large fraction of 
the neutralino relic abundance falls into the restricted range 
$0.02 \lsim \Omega_{\rm CDM} h^2 \lsim 0.2$, which turns out to be the most
appealing interval for relic neutralinos as indicated from recent observations
and analyses on the value of the matter content of the Universe \cite{mssm_mod}.

The properties of set $S$ relevant to searches at accelerators
are displayed in Fig.3. Section (a) of this figure 
shows a scatter plot of set $S$ in term  of $m_h$ and $\tan \beta$,
where it is apparent  a correlation between $\tan \beta$ 
and $m_h$. This is due to the fact that the rather 
large values
$\sigma_{\rm scalar}^{(\rm nucleon)} \sim (10^{-9} - 10^{-8}$) nb, as required
 by the annual modulation data, impose that either the couplings are large 
(then large $\tan \beta$) and/or the
process goes through the exchange of  a light particle. 
Thus, Higgs--exchange dominance (which turns out to occur here) and 
$\sigma_{\rm scalar}^{(\rm nucleon)} \sim (10^{-9} - 10^{-8})$ nb require a 
very light $h$ at small $\tan \beta$, 
and even  put {\it a lower bound on tan $\beta$: $\tan \beta \gsim 2.5$}. 
{From} Fig.3(a) we notice that a good deal of susy
configurations are explorable at LEP2, while others will require experimental
investigation at a high luminosity Fermilab Tevatron, which 
should be capable to explore Higgs masses up to $m_h \sim$ 130 GeV. 
In Fig.3(b) we display the scatter plot of set $S$ in the plane 
$m_{\chi}$ -- $\tan \beta$. Since the reach of LEP2 extends only up to
the dashed vertical line, at $m_\chi \simeq 50$ GeV, the exploration of the 
whole interesting region will require Tevatron upgrades or LHC. Under favorable
hypothesis, TeV33 could provide exploration up to the vertical solid line. 

\vspace{-5.0mm}
\section{Analysis in SUGRA schemes}

\begin{figure}[t]
\hbox{
\hspace{-3.5mm}
\psfig{figure=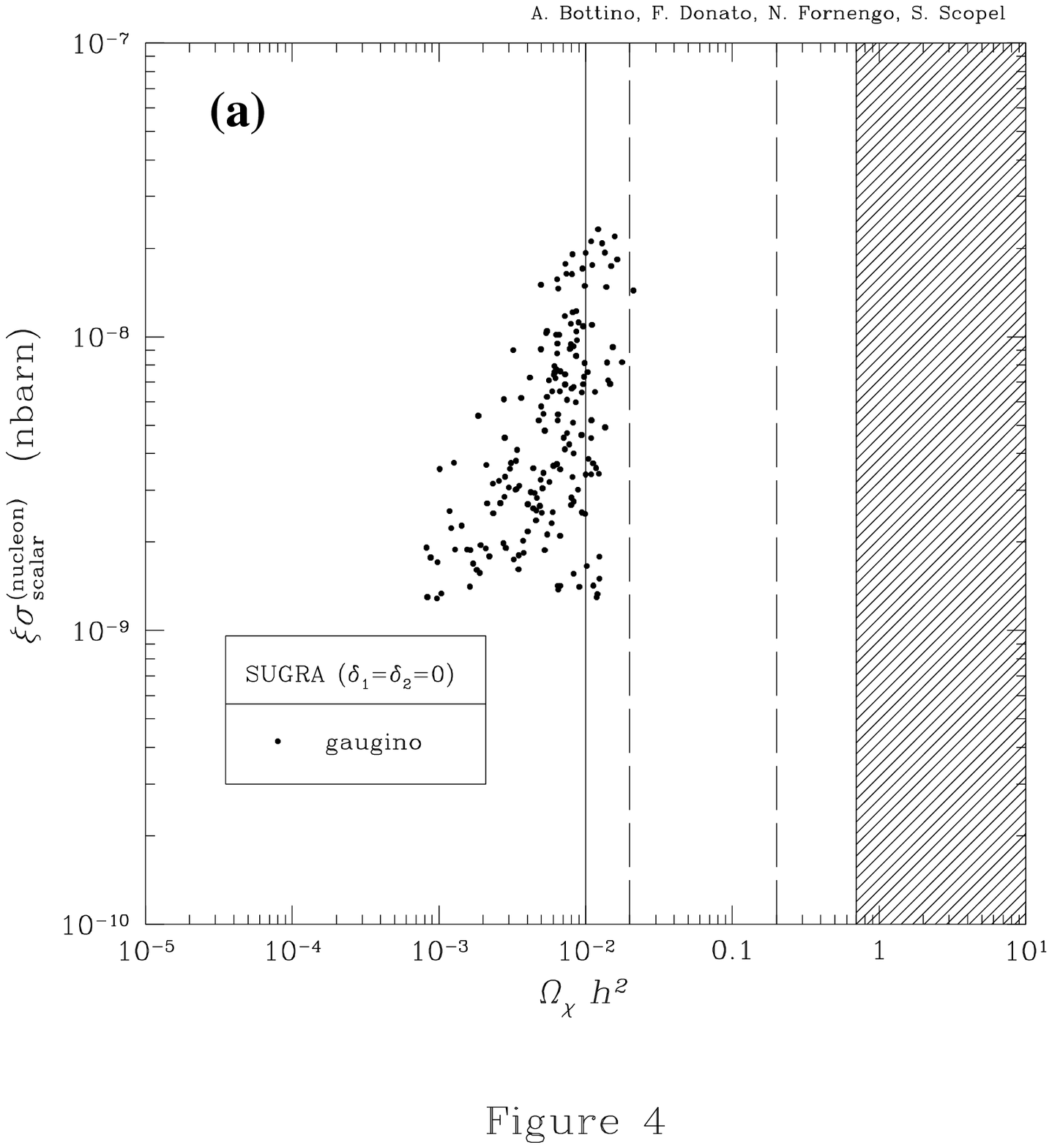,width=2.45in,bbllx=60bp,bblly=222bp,bburx=530bp,bbury=675bp,clip=}
\hspace{-4.5mm}
\psfig{figure=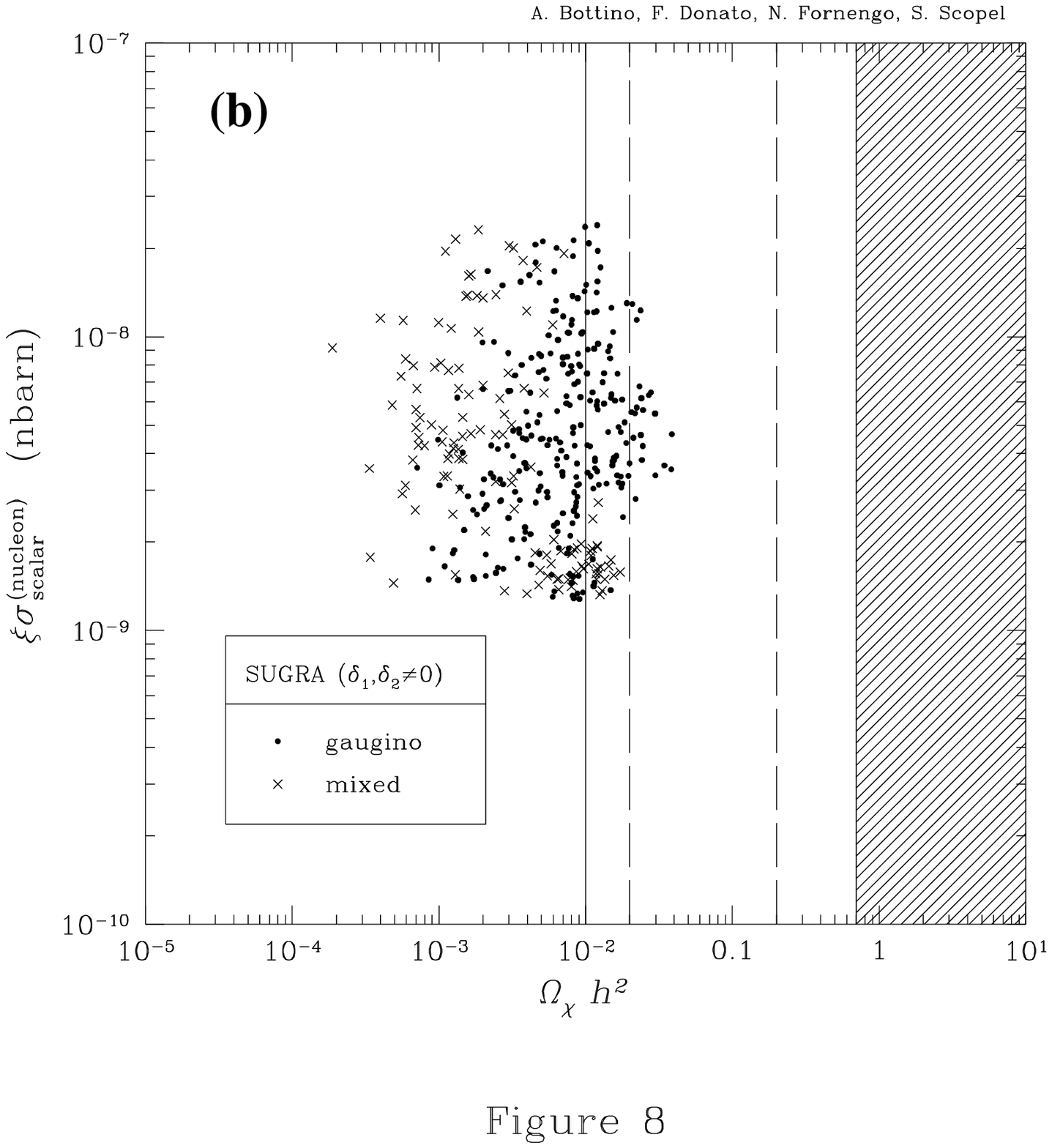,width=2.45in,bbllx=60bp,bblly=222bp,bburx=530bp,bbury=675bp,clip=}
}
\vspace{-3.0mm}
\caption{Scatter plot of {\bf SUGRA} configurations compatible with the annual
modulation data in the plane 
 $\Omega_{\chi} h^2$ -- $\xi \sigma_{\rm scalar}^{(\rm nucleon)}$. 
(a): universal SUGRA models. (b): SUGRA models with
deviations from universality in the Higgs sector.
Notations are as in Fig.2.}
\vspace{-4.0mm}
\end{figure}

\begin{figure}[t]
\hbox{
\hspace{-3.5mm}
\psfig{figure=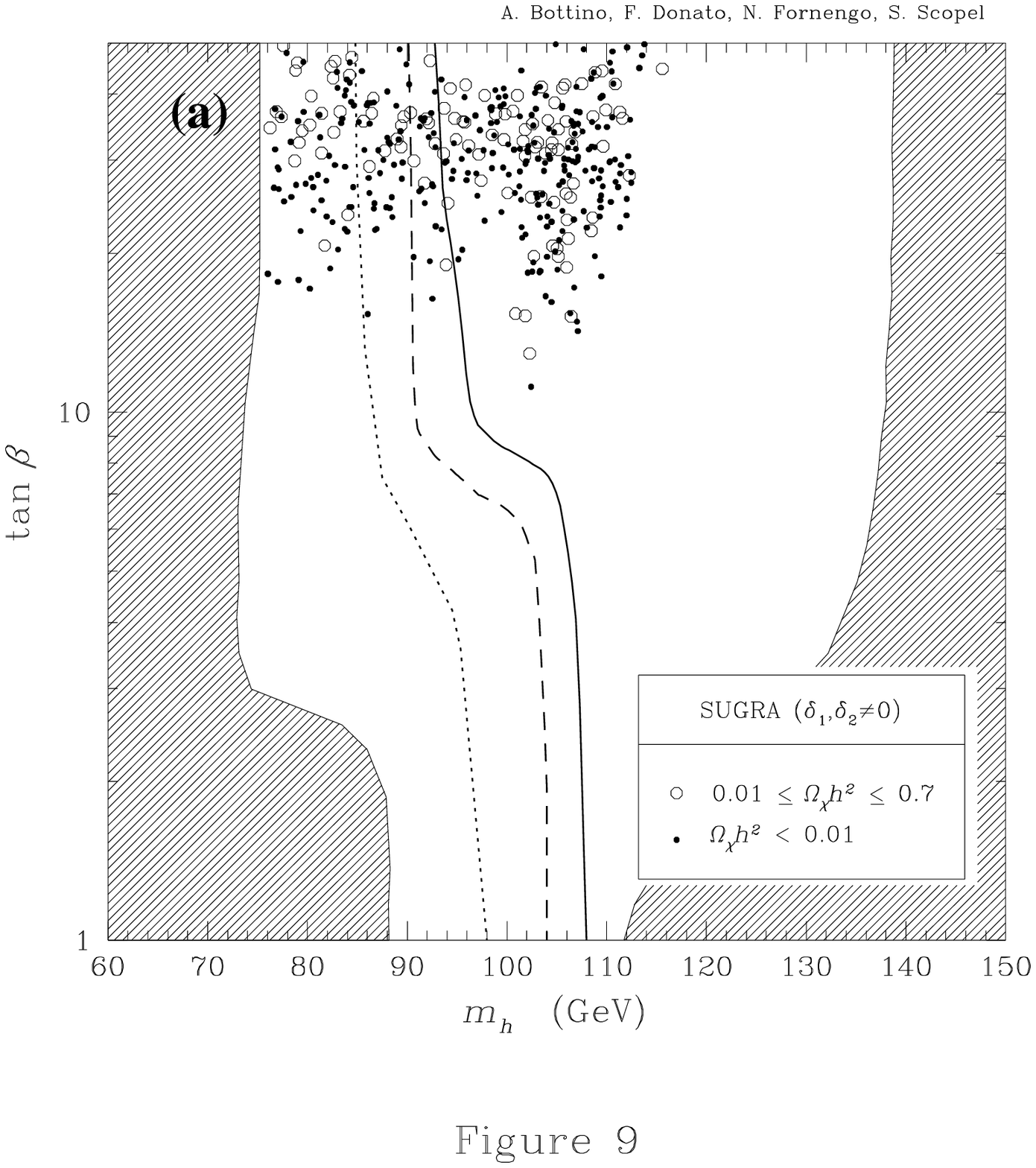,width=2.45in,bbllx=60bp,bblly=222bp,bburx=530bp,bbury=675bp,clip=}
\hspace{-4.5mm}
\psfig{figure=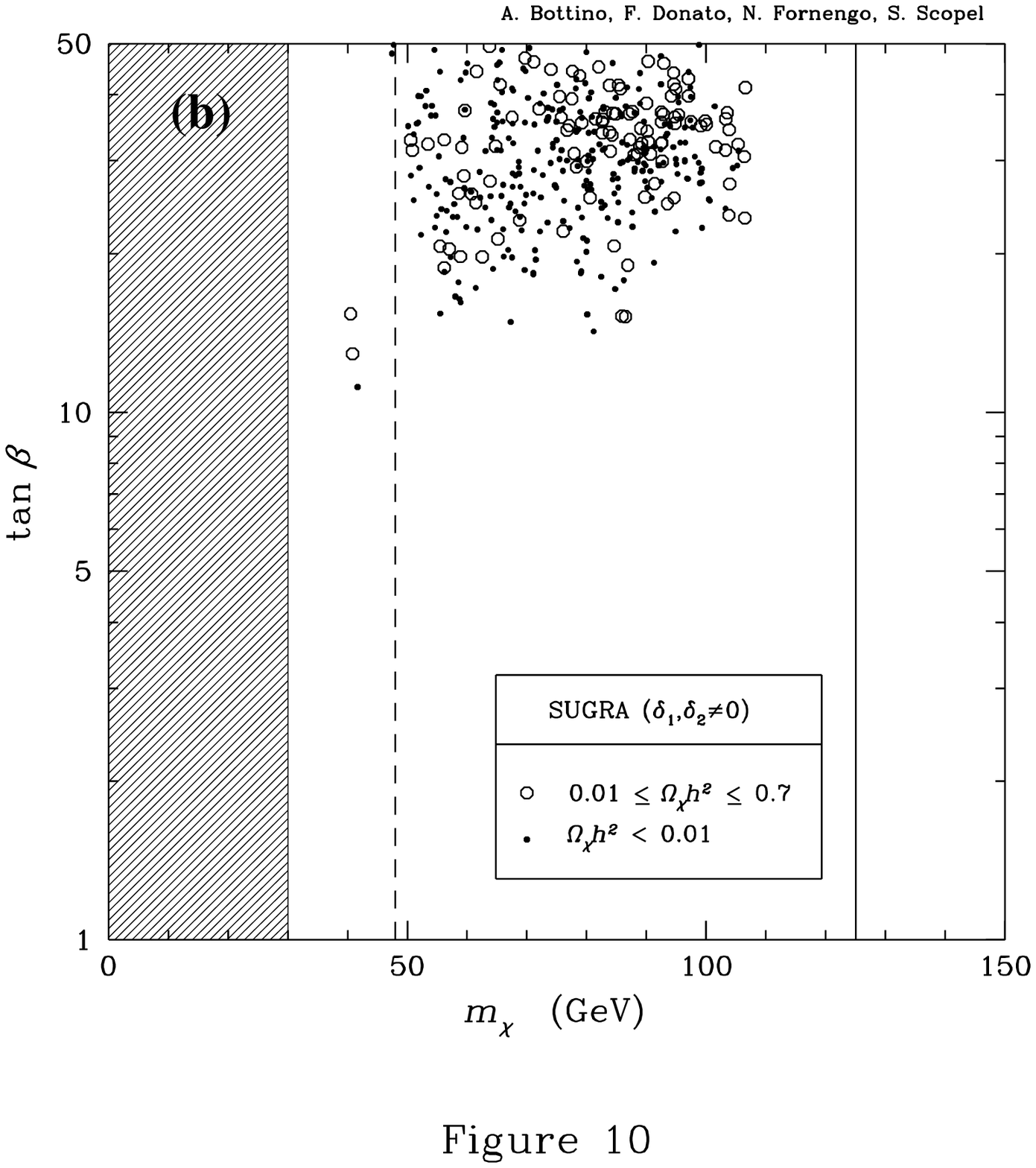,width=2.45in,bbllx=60bp,bblly=222bp,bburx=530bp,bbury=675bp,clip=}
}
\vspace{-3.0mm}
\caption{Explorability at accelerators of the {\bf non--universal SUGRA} configurations 
compatible with the annual modulation data. (a): Scatter plot in the plane 
$m_h$ -- $\tan \beta$. (b): Scatter plot in the plane $m_{\chi}$ -- $\tan \beta$. 
Notations are as in Fig.3.}
\vspace{-4.0mm}
\end{figure}

\vspace{-3.0mm}
In this Section we show that the susy features,
implied by the DAMA/NaI data, are also compatible with more 
ambitious supersymmetry 
schemes, where the previous phenomenological model is implemented in 
a supergravity (SUGRA) framework, especially if the unification 
conditions, which are frequently imposed at the Grand  
Unification (GUT) scale, are appropriately relaxed \cite{sugra_mod}. 

The essential elements of the SUGRA models employed here are:
a Yang--Mills Lagrangian, the
superpotential and  the soft--breaking
Lagrangian. In this class
of models the electroweak symmetry breaking (EWSB) is induced radiatively.
This supergravity  framework  is usually implemented by
some restrictive assumptions about unification at
$M_{GUT}$: (i) unification  of the gaugino masses: $M_i(M_{GUT}) \equiv m_{1/2}$, 
(ii) universality of the scalar masses: $m_i(M_{GUT}) \equiv m_0$,
(iii) universality of the trilinear scalar couplings:
$A^{l}(M_{GUT}) = A^{d}(M_{GUT}) = A^{u}(M_{GUT}) \equiv A_0 m_0$. 
As extensively discussed in Ref. \cite{bere}, these conditions have strong 
consequences for low--energy supersymmetry
phenomenology, and in particular for the properties of the neutralino.

The unification conditions represent a theoretically attractive
possibility, which makes strictly universal SUGRA models very predictive.
However, the above assumptions, particularly (ii) and (iii), are not fully justified, 
since universality may occur at a scale higher
than $M_{GUT}$, i.e. the Planck scale or string scale, 
in which case renormalization above $M_{GUT}$ weakens universality.
We therefore discuss the DAMA/NaI data both in a SUGRA model 
with strict unification conditions and in a SUGRA framework, where 
we introduce a departure from universality in the scalar masses at $M_{GUT}$ 
which splits the Higgs mass parameters:
$ M^2_{H_i}(M_{GUT}) = m_0^2(1+\delta_i)$.
The parameters $\delta_i$ will be varied in the range ($-1$,$+1$), but are taken to be
independent of the other susy parameters. 

Because of the requirements of radiative EWSB and
of the universality conditions, the independent susy parameters are reduced to 
(apart from the $\delta_i$'s): $m_{1/2}, m_0, A_0, \tan \beta$
and ${\rm sign}(\mu)$. They are varied in the following ranges:
$10\;\mbox{GeV} \leq m_{1/2} \leq  500\;\mbox{GeV},\;
 m_0 \leq  1\;\mbox{TeV},\;
-3 \leq A_0 \leq +3,\;
1 \leq \tan \beta \leq 50$; the parameter $\mu$ is taken positive.
The values taken as upper limits of the ranges for 
$m_{1/2}, m_0$ are inspired by the upper 
bounds which may be
derived for these quantities in SUGRA theories, when one requires that the 
EWSB, radiatively induced by the soft supersymmetry
breaking, does not occur with excessive fine tuning. The same argument
was also used in the previous Section in setting the upper limits on 
the dimensional parameters of the MSSM. 

The susy parameter space is constrained by the same experimental
bounds discussed in the previous Section for the MSSM, with the additional
constraint arising from the limits on the bottom--quark mass $m_b$. The 
bottom mass is computed as a function of the susy
parameters  and required to be compatible with the present 
experimental bounds \cite{mb}.

The susy configurations compatible with the
DAMA data are shown in the plane 
$\xi \sigma^{(\rm nucleon)}_{\rm scalar}$ -- $\Omega_\chi h^2$
in Fig.4(a) for universal SUGRA models and in Fig.4(b) for
models with deviation from strict universality.
We notice that also in SUGRA theories a fraction of the selected susy
configurations fall into the cosmologically interesting
range of $\Omega_\chi h^2$.

Other qualifications for the configurations which lie inside the
region $R$, which are relevant for
searches at accelerators, concern the ranges for the $h$--Higgs boson mass, 
the neutralino mass and the lightest top--squark mass. In the case
of universal SUGRA models, we find:
$m_h \lsim 115$ GeV, 50 GeV $\lsim m_{\chi} \lsim$ 100 GeV,
200 GeV $\lsim m_{\tilde t_1} \lsim$ 700 GeV and 
$\tan\beta \gsim 42$. For deviations from universality, the situation
is shown in Fig.5 where we notice that the sample of representative 
points covers a slightly wider domain.
The ranges of the Higgs and neutralino masses
are similar to those already found in the universal case,
but now $\tan\beta$ extends to the interval $10 \lsim \tan \beta \lsim 50$,
instead of being limited only to very large values.

\vspace{-5.0mm}
\section{Indirect detection of neutralino dark matter}

\begin{figure}[t!]
\centerline{
\psfig{figure=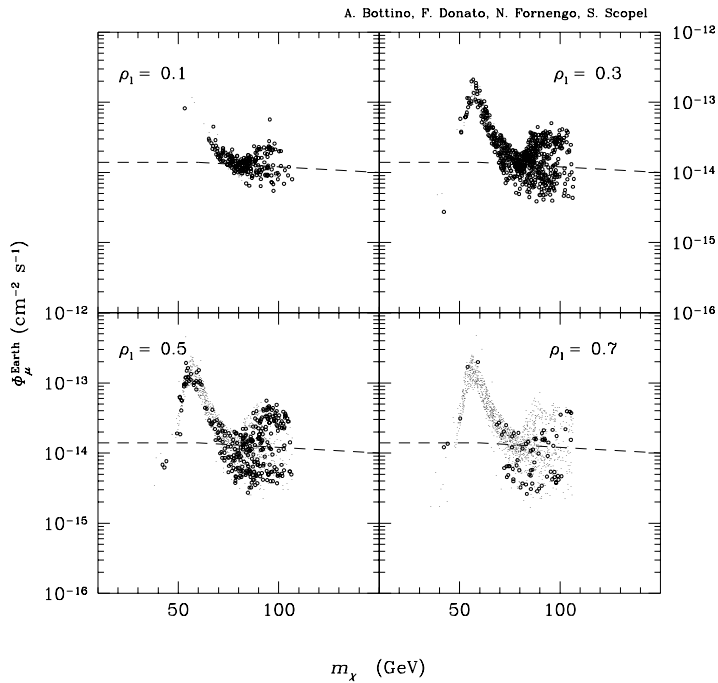,width=3.90in,bbllx=40bp,bblly=180bp,bburx=550bp,bbury=650bp,clip=}
}
\vspace{-3.0mm}
\caption{Scatter plots for the up--going muon fluxes from the center of the Earth versus
the neutralino mass. The {\bf MSSM} configurations compatible with the annual modulation data are
subdivided into the 4 panels, depending on the corresponding value 
the local density:
$\rho_l/({\rm GeV cm}^{-3})$ = $0.1$, $0.3$, 
$0.5$, $0.7$. 
Dots denote configurations which could be excluded on the basis of the BESS95 
antiproton data, circles denote configurations which survive this exclusion
criterion. The dashed line denotes the MACRO upper bound. 
}
\vspace{-4.0mm}
\end{figure}

\vspace{-3.0mm}
The susy configurations which have been proved to
be compatible with the annual modulation data can also
be searched for by using methods of indirect 
search for relic particles \cite{ind_mod}. The two most promising techniques
are the measurement of cosmic--ray antiprotons \cite{pierre} and the
measurement of neutrino fluxes from Earth and Sun \cite{indiretta}.

The signals we are going to discuss here
consists of the fluxes of up--going muons in a neutrino telescope, generated by
neutrinos which are produced by pair annihilations
of neutralinos captured and accumulated inside the Earth and the
Sun. The calculation of the up--going muon signal is performed according to the
method described in Ref. \cite{indiretta}.

In Fig. 6 we display the 
scatter plots for the flux of the up--going muons from the center of the 
Earth, for various values of the  local dark matter density 
$\rho_l$. In this figure is also reported the current $90\%$ C.L. 
experimental upper 
bound on $\Phi_{\mu}^{\rm Earth}$, obtained by MACRO \cite{macro}.
We notice the particular enhancement at 
$m_{\chi} \sim (50 - 60)$ GeV, due to a mass--matching effect
between the WIMP and the Fe nuclei in the Earth core.
It is also noticeable an effect of suppression and spreading 
of the fluxes for $m_{\chi} \gsim 70$ GeV at  
$\rho_l \gsim 0.3$ GeV cm$^{-3}$. 
This is due to the fact that the configurations with these 
large values of $\rho_l$ may imply, because of the annual modulation data,
a neutralino--nucleus cross--section 
which is too small to establish an efficient capture rate, necessary for 
the capture-annihilation 
equilibrium in Earth \cite{ind_mod}.
In Fig. 6  the configurations which would be 
excluded on the basis of the antiproton data 
are denoted differently from those which would survive this 
criterion (for details see Refs. \cite{ind_mod,fiorenza}).

By comparing our scatter plots with the experimental MACRO upper limit, 
one notices that a number of susy configurations provide 
a flux in excess of this experimental bound and might then be considered
as excluded. However, 
it has to be recalled that a possible neutrino oscillation 
effect may 
be operative here and  affect   the indirect neutralino signal as 
well as the background consisting in atmospheric neutrinos.  
Therefore a strict enforcement of the
current upper bound on $\Phi_\mu^{\rm Earth}$ should be applied with caution
as long as the neutrino oscillation properties are not fully considered.  
However, it is rewarding that the set $S$ of 
susy configurations is quite accessible to relic neutralino 
indirect search by measurements of up--going fluxes.

\vspace{-5.0mm}
\section{Conclusions}

\vspace{-3.0mm}
In this paper, we have analysed in terms of relic neutralinos
the total sample of new 
and former DAMA/NaI data \cite{dama1,dama2}, which provide 
the indication of a possible annual modulation effect in the 
rate for WIMP direct detection.
The remarkable result of our analysis is that {\em the annual
modulation data are widely compatible with a relic neutralino making 
up the major part of dark matter in the Universe}, both in the low
energy MSSM and in SUGRA schemes.

We have also investigated the possibility of exploring at accelerators
the same susy configurations which are compatible with the 
annual modulation data. We have shown that an analysis of the main 
features of these susy configurations is within the reach of present 
or planned experimental set--ups. 
In particular, we have obtained the following results:

\vspace{-2.0mm}
\begin{description}
\item[$\bullet$]
      {\underline{MSSM:}}
      The sizeable neutralino--nucleon elastic cross--sections, implied by the 
      annual modulation data, entail  a rather stringent upper bound for 
      $m_h$ in terms of $\tan \beta$. In particular, this property  implies that no 
      susy configuration would be allowed for $\tan \beta \lsim 2.5$. 
      Another property, discussed in Ref. \cite{mssm_mod}, is that
      the annual modulation data and the $b \rightarrow s + \gamma$ 
      constraint complement each other in providing a correlation 
      between $\tan \beta$ and the mass of the lightest top--squark. 
\item[$\bullet$]
      \vspace{-2.0mm}
      {\underline{SUGRA:}}
      In the universal SUGRA model the constraints imposed by the DAMA/NaI data 
      imply for the $h$--Higgs boson mass, 
      the neutralino mass and the lightest top--squark mass, the following ranges: 
      $m_h \lsim 115$ GeV, 50 GeV $\lsim m_{\chi} \lsim$ 100 GeV and 
      200 GeV $\lsim m_{\tilde t_1} \lsim$ 700 GeV, respectively. 
      In universal SUGRA $\tan \beta$ is constrained to be large, 
      $\tan \beta \gsim 42$, whereas, 
      with departure from universality in the scalar masses, the range for 
      $\tan \beta$ widens to $10 \lsim \tan \beta \lsim 50$. 
\end{description}

\vspace{-2.0mm}
Many of the above configurations will be explored by LEP2, and almost all
of them are under reach of the future planned high--energy accelerators, namely
the upgrade of the Tevatron and LHC.

The same configurations can also be probed by indirect dark matter searches.
We have shown \cite{ind_mod} that a sizeable fraction of the 
susy neutralino configurations singled out by the 
DAMA/NaI data may provide signals detectable by
measurement of cosmic--ray antiprotons and detection of 
neutrino fluxes from Earth and Sun. 

For the case of cosmic antiprotons, it has been shown \cite{ind_mod,fiorenza}
that present data are well fitted by total spectra which include a $\bar p$
contribution from neutralino--pair annihilation, with neutralino 
configurations which are relevant for annual modulation in direct detection. 
These data can also be used to reduce the total sample of the
susy configurations under study, and to narrow the range 
of the local density, by disfavoring its largest values.
Investigation by measurements of cosmic $\bar p$'s looks very promising 
in view of the collections and analyses of more statistically 
significant sets of data in the 
low--energy regime which are currently under way and
which may soon provide further relevant information, like, for instance, 
in the case of BESS and AMS detectors.

Measurements of neutrino fluxes from Earth and Sun, due to capture and annihilation
of neutralinos inside these celestial bodies, have been proved to be sensitive
to neutralino configurations singled out by the annual modulation data. However, an
appropriate interpretation of these measurements preliminarily requires 
some clarification of the oscillation neutrino properties, especially in the
light of the recent Kamiokande result \cite{kamioka}.

Finally, we conclude with a few comments. A solid confirmation of the
annual modulation effect, as singled out by the DAMA/NaI Collaboration,
necessarily requires further accumulation of data with very stable 
set--ups over a few years, a project which is currently
undertaken by the DAMA/NaI Collaboration. For the status of the
experimental activity in WIMP direct search with other detectors,
see the contributions of the various experimental Collaborations to
these Proceedings.

It is really worth noticing that the detection of the effect of annual
modulation, if confirmed by further experimental evidence, would 
turn out to be a major breakthrough in establishing the existence of 
particle dark matter in the Universe, and this would henceforth be a 
major breakthrough for astrophysics, cosmology and particle physics as well.

\end{document}